\documentstyle[12pt]{article}
\newcommand{\z}{&&\hspace*{-1cm}}
\begin{document}

\setcounter{page}{0}
\thispagestyle{empty}

{}~\vspace{2cm}

\begin{center} {\Large {\bf SMALL $X$ BEHAVIOUR OF PARTON
      DISTRIBUTIONS IN PROTON}} \\
 \vspace{1.5cm}
 {\large
  A.V.Kotikov\footnote{
On leave of absence from Particle Physics Laboratory, JINR, Dubna, Russia.\\
e-mail: KOTIKOV@LAPPHP0.IN2P3.FR, KOTIKOV@SUNSE.JINR.DUBNA.SU} \\
 \vspace {0.5cm}
 {\em Laboratoire Physique Theorique ENSLAPP\\
LAPP, B.P. 110, F-74941, Annecy-le-Vieux Cedex, France}}
% {\em Laboratory of Particle Physics,
%  Joint Institute for Nuclear Research,
% 141980 Dubna (Moscow Region), Russia}}
\end{center}

\vspace{1.5cm}\noindent
\begin{center} {\bf Abstract} \end{center}

 The paper presents the $QCD$ description of the hard and semihard
 processes in the framework of the Wilson operator product expansion.
The smooth transition
 between the cases of the soft and hard Pomerons is obtained.
%Some analysis is given for new {\it HERA} data .

%PACS number(s): 13.60.Hb, 12.38.Bx, 13.15.Dk
\newpage
\pagestyle{plain}

The recent measurements of the deep-inelastic (DIS) structure function
(SF) $F_2$ by the $H1$ \cite{1} and $ZEUS$ \cite{2} collaborations
open a new kinematical range to study proton structure. The new
$HERA$ data show the strong increase of $F_2$ with decresing
$x$. However, the data of the $NMC$ \cite{3} and $E665$ collaboration
\cite{4} at small $x$ and smaller $Q^2$ is in the good agreement with
the standard Pomeron or with the Donnachie-Landshoff picture where the
Pomeron intercept: $\alpha_p = 1.08$, is very close to standard
one. The interpritation of the fast changing of the intercept in
the region of $Q^2$ between $Q^2=1GeV^2$ and $Q^2=10GeV^2$ (see Fig.3
in \cite{5}) is yet absent. There are the
arguments in favour of that is one intercept (see \cite{6}) or the
superposition of two different Pomeron trajectories, one having an
intercept of $1.08$ and the one of $1.5$ (see Fig.4 in \cite{5}).

The aim of this article is the possible ``solution'' of this problem in
the framework of
%standard
Dokshitzer-Gribov-Lipatov-Altarelli-Parisi (DGLAP) equation \cite{6.5}. It is
good known (see, for example, \cite{7}), that in the
double-logarithmical approximation the DGLAP equation solution is the
Bessel function, or $\exp{\sqrt{\phi (Q^2) ln(1/x)}}$, where
$\phi (Q^2)$ is known $Q^2$-dependent function\footnote{More
  correctly, $\phi$ is $Q^2$-dependent for the solution of DGLAP
equation with the  boundary condition: $f_a(x,Q^2_0)= Const$ at
  $x \to 0$. In the case of the boundary condition: $f_a(x,Q^2_0) \sim
  \exp{\sqrt{ ln(1/x)}}$, $\phi$ is lost (see \cite{8}) its
  $Q^2$-dependence}. However, we will seek the
``solution''\footnote{We use
  the termin ``solution'' because we will work in the leading twist
  approximation in the range of $Q^2$: $Q^2>1GeV^2$, where the higher
  twist terms may give the sizeable contribution (see, for example,
  \cite{9}). Moreover, our ``solution'' is the Regge asymptotic with
  unknown parameters rather then the solution of DGLAP equation. The
  parameters are found from the agreement of the r.h.s. and l.h.s. of
  the equation.}
 of DGLAP
equation in the Regge form (we use the parton distributions (PD)
multiplied by $x$ and neglect the nonsinglet quark distribution at
small $x$):
 \begin{eqnarray}
f_a (x,Q^2) \sim x^{-\delta} \tilde
f_a(x,Q^2), ~~~~(a=q,g)~~(\alpha_p \equiv 1+\delta ) \label{1}
 \end{eqnarray}
 where $\tilde f_a(x,Q^2)$ is nonsingular at $x \to 0$ and $\tilde
f_a(x,Q^2) \sim (1-x)^{\nu}$ at $x \to 1$\footnote{Consideration of
  the more complicate behaviour in the form $x^{-\delta}(ln(1/x))^b
  I_{2g}(\sqrt{\phi ln(1/x)})$ is given in \cite{8} and will be
  considered in this content in the forthcomming article \cite{10}}. The
similar investigations were already done and the results are good
known (see  \cite{8}, \cite{11}-\cite{15})\footnote{In the
  double-logarithmical approximation the similar results were obtained
  in \cite{15.5}}. The aim of this letter is to expand
these results to the range where $\delta \sim 0$ (and $Q^2$ is not
large) following to the observed early (see \cite{12,13})\footnote{The
method is based on the earlier results \cite{16}}
method to replace the Mellin convolution by a simple product. Of
course, we understand that the Regge behaviour (\ref{1}) is not in
the agreement with the double-logarithmic solution, however the range,
where $\delta \sim 0$ and the $Q^2$ values are nonlarge, is really the
Regge regime and a ``solution''
%(\ref{1})
of DGLAP equation in the form of (\ref{1})
would be worthwhile. This ``solution'' may be understand as the
solution of DGLAP equation together with the condition of its Regge
asymptotic at $x \to 0$.

Consider DGLAP equation and apply the method from \cite{13} to the
Mellin convolution in its r.h.s. (in contrast with standard case, we
use below $\alpha(Q^2)=\alpha_s(Q^2)/(4\pi)$):
 \begin{eqnarray}
\z \frac{d}{dt}f_a (x,t)~=~- \frac{1}{2} \sum_{i=a,b}
\hat \gamma_{ai}(\alpha,x) \otimes f_a(x,t)~~~(a,b)=(q,g) \nonumber \\
\z =~- \frac{1}{2} \sum_{i=a,b}
\tilde \gamma_{ai}(\alpha,1+\delta)f_a(x,t)~+~O(x^{1-\delta})~~~
\Bigl( \gamma_{ab}(\alpha,n)=\alpha \gamma_{ab}^{(0)}(n)+\alpha^2
\gamma_{ab}^{(1)}(n)+...\Bigr)
, \label{2}
 \end{eqnarray}
where $t=ln(Q^2/\Lambda ^2)$. The $\hat \gamma_{ab}(\alpha,x)$ are the
spliting functions corresponding to the anomalous dimensions (AD)
$\gamma_{ab}(\alpha,n) = \int^1_0 dx x^{n-1} \hat \gamma_{ab}(\alpha,x)$. Here
the functions $\gamma_{ab}(\alpha,1+\delta)$
are the AD $\gamma_{ab}(\alpha,n)$ expanded from the
integer argument ``$n$'' to the noninteger one ``$1+\delta$''. The
functions $\tilde \gamma_{ab}(\alpha,1+\delta)$ (marked lower as AD, too)
can be obtained from
the functions $\gamma_{ab}(\alpha,1+\delta)$  replacing the term $1/\delta$
by the one $1/\tilde \delta$:

 \begin{eqnarray}
\frac{1}{\delta} \to \frac{1}{\tilde \delta}~=~\frac{1}{\delta}
\Bigl( 1 - \varphi(x,\delta)x^{\delta} \Bigr)
 \label{3}
 \end{eqnarray}
%if, of course, the term $1/\delta$ is contained into
%$\gamma_{ab}(\alpha,1+\delta)$.
This replacement (\ref{3}) is appeared
very naturally from the consideration the Mellin convolution at $x \to
0$ (see \cite{13}) and preserves the smooth and nonsingular transition to
the case $\delta =0$, where

 \begin{eqnarray}
 \frac{1}{\tilde \delta}~=~ln\frac{1}{x} - \varrho(x)
 \label{4}
 \end{eqnarray}

The concrete form of the functions $\varphi(x,\delta)$ and
$\varrho(x)$ depends strongly on the type of the behaviour of the
PD $f_a(x,Q^2)$ at $x \to 0$ and in the case of the Regge regime (\ref{1})
they are (see \cite{12,13}):

 \begin{eqnarray}
\varphi(x,\delta)~=~ \frac{\Gamma(\nu +1)\Gamma(1-\delta)}{\Gamma(\nu
  +1-\delta)}~~ \mbox{ and }~~ \varrho(x)~=~\Psi(\nu+1)-\Psi(1),
 \label{5}
 \end{eqnarray}
where $\Gamma(\nu+1)$ and $\Psi(\nu+1)$ are the Eulerian $\Gamma$- and
$\Psi$-functions, respectively. As it can be seen, there is the
correlation with the PD behaviour at large $x$.

If $\delta$ is not small (i.e. $x^{\delta}>>1$), we can replace
$1/ \tilde \delta$ to $1/\delta$ in the r.h.s. of Eq.(\ref{2}) and
obtain its solution in the form (hereafter $t_0=t(Q^2=Q^2_0)$):

 \begin{eqnarray}
 \frac{f_a(x,t)}{f_a(x,t_0)}~=~ \frac{M_a(1+\delta,t)}{M_a(1+\delta,t_0)},
 \label{6}
 \end{eqnarray}
 where $M_a(1+\delta,t)$ is the analytical expansion of the PD moments
 $M_a(n,t) = \int^1_0 dx x^{n-1} f_a(x,t)$ to the noninteger
value ``$n=1+\delta$''.

This solution is good known one (see \cite{12} for the first two
orders of the perturbation theory, \cite{14} for the first three
orders and \cite{15} containing a resummation of all orders,
respectively). Note that recently the fit of $HERA$ data was
done in \cite{17} with the formula for PD $f_q(x,t)$ very close
\footnote{The used  formula (Eq.(2) from \cite{17}) coincides with
  (\ref{6}) in the leading order (LO) approximation, if we save
only $f_g(x,Q^2)$ in the
  r.h.s. of (\ref{2}) (or put $\gamma_{qq}=0$ and $\gamma_{qg}=0$
  formally). Eq.(\ref{6}) and Eq.(\ref{2}) from \cite{17} have
    some differences in the next-to-leading order (NLO), which are not
very important because they are
    corrections to the $\alpha$-correction.} to
(\ref{6}) and the very well agreement (the $\chi^2$ per degree of freedom
is $0.85$) is found at $\delta = 0.40 \pm 0.03$. There are also the
fits \cite{17.5} of the another group using equations which are
similar to (\ref{6}) in the LO approximation.

The news in our investigations are in the follows. Note that the
$Q^2$-evolution of $M_a(1+\delta,t)$ contains  the two: ``+'' and
``$-$'' components, i.e. $M_a(1+\delta,t)= \sum_{i= \pm} M_a^i(1+\delta,t)$,
and in principle the every component evolves separately and may have
the independent (and not equal) intercept. Here for the
simplicity we restricte ourselves to the LO analysis and give NLO
formulae lower without large intermediate equations.

{\bf 1.} Consider DGLAP equation for the ``+'' and ``$-$'' parts
(hereafter $s=ln(lnt/lnt_0)$):

  \begin{eqnarray}
 \frac{d}{ds} f_a^{\pm}(x,t)~=~- \frac{1}{2\beta_0}
\tilde
\gamma_{\pm}(\alpha,1+\delta_{\pm})f_a^{\pm}(x,t)~+~O(x^{1-\delta}),
 \label{7} \end{eqnarray}
where
$$\gamma_{\pm}~=~ \frac{1}{2}
\biggl[
\Bigl(\gamma_{gg}+\gamma_{qq} \Bigr)~\pm ~
\sqrt{ {\Bigl( \gamma_{gg}- \gamma_{qq} \Bigr)}^2
  ~+~4\gamma_{qg}\gamma_{gq}}
\biggr]$$
are the AD of the ``$\pm $'' components (see, for example, \cite{18})

The ``$-$'' component $\tilde \gamma_{-}(\alpha,1+\delta_-)$ does not
contain the singular term (see \cite{12,14} and lower) and its
solution
have the form:

 \begin{eqnarray}
 \frac{f_a^-(x,t)}{f_a^-(x,t_0)}~=~e^{-d_-(1+\delta_-)s}, \mbox{ where }
d_{\pm}=\frac{\gamma_{\pm}(1+\delta_{\pm})}{2\beta_0}
 \label{8}
 \end{eqnarray}

The ``+'' component $\tilde \gamma_{+}(\alpha,1+\delta_+)$
contains the singular term  and $f_a^+(x,t)$  have the solution similar
(\ref{8})
%(but with the replace ``$-$'' to ``+'')
only for $x^{\delta_+}>>1$:

 \begin{eqnarray}
 \frac{f_a^+(x,t)}{f_a^+(x,t_0)}~=~e^{-d_+(1+\delta_+)s}, \mbox{ if }
%d_{\pm}=\frac{\gamma_{\pm}(1+\delta_{\pm})}{2\beta_0}
 x^{\delta_+}>>1
 \label{9}
 \end{eqnarray}

 The both intersepts $1+\delta_+$ and $1+\delta_-$ are
unknown  and should be found, in principle, from the analysis of the
experimental
data. However there is the another way. From the small $Q^2$ (and
small $x$) data of the $NMC$ \cite{3} and $E665$ collaboration \cite{4} we
can conclude that the SF $F_2$ and hence the PD $f_a(x,Q^2)$ have the
flat asymptotics for $x \to 0$ and $Q^2 \sim (1\div2)GeV^2$. Thus we
know that the values of $\delta_+$ and $\delta_-$ is approximately
zero at $Q^2 \sim 1GeV^2$.

 Consider  Eqs.(\ref{7}) with
$\delta_{\pm}=0$ and with the boundary condition $f_a(x,Q^2_0)=A_a$ at
$Q^2_0=1GeV^2$. For the ``$-$'' component we already have the solution:
the Eq.(\ref{8}) with $\delta_-=0$ and $d_-(1)=16f/(27\beta_0$), where
$f$ is the number of the active quarks and $\beta_i$ are the
coefficients in the $\alpha$-expansion of QCD $\beta$-function.
For its ``+'' component
Eq.(\ref{7}) can be rewritten in the form (hereafter the index
$1+\delta $ will be omitted in the case $\delta \to 0$):

  \begin{eqnarray} ln(\frac{1}{x})\frac{d}{ds}\delta_+(s)~+~
 \frac{d}{ds} ln(A_a^+) ~=~- \frac{1}{2\beta_0}
\biggr[ \hat
\gamma_{+}
\Bigl( ln(\frac{1}{x}) -\varrho(\nu)  \Bigr) ~+~ \overline \gamma_+
\biggl]
 \label{10} \end{eqnarray}
where
$\hat\gamma_{+}$ and $\overline \gamma_+$ are the coefficients of the
singular and regular parts at $\delta \to 0$ of AD
$\gamma^+(1+\delta)$:
$$ \gamma^+(1+\delta)~=~\hat\gamma^+ \frac{1}{\delta} ~+~
\overline\gamma^+,~~~\hat\gamma^+=-24,~\overline\gamma^+=22+
\frac{4f}{27}$$

The solution of Eq.({10}) is

 \begin{eqnarray}
 f_a^+(x,t)~=~A^+_a~x^{\hat d_+s}e^{-\overline d_+s},
 \label{11}
 \end{eqnarray}
where
$$\hat d_+ \equiv \frac{\hat \gamma^+}{2\beta_0} \simeq -
\frac{4}{3},~~
\overline d_+ \equiv \frac{1}{2\beta_0}
\Bigl( \overline \gamma_+  ~-~ \hat \gamma_+ \varrho(\nu)\Bigr)
 \simeq \frac{4}{3} \varrho(\nu) + \frac{101}{81}$$
Herefter the symbol $\simeq $ marks the case $f=3$.

As it can be seen from (\ref{11}) the flat form $\delta_+=0$ of the
``+''-component of PD is very nonstable from the (perturbative)
viewpoint, because $d(\delta_+)/ds \neq 0$, and for $Q^2 > Q_0^2$ we
have already the nonzero power of $x$ (i.e. pomeron intercept
$\alpha_p >1$). This is in the agreement with the experimental data. Let us
note that the power of x is positive for $Q^2<Q^2_0$ that is in principle
also supported by the $NMC$ \cite{3}
%and $E665$ callaboration \cite{4}
data, but the use of this analysis to $Q^2<1GeV^2$ is open the
question.

Thus, we have the DGLAP equation solution for the ``+'' component at $Q^2$
is close to $Q^2_0=1GeV^2$, where Pomeron starts in its movement to the
subcritical (or Lipatov \cite{19.5,19.6}) regime and also for the large
$Q^2$, where pomeron have
the  $Q^2$-independent intercept. In principle, the general
solution of (\ref{7}) should contain the smooth transition between
these pictures but this  solution is absent
%(at least, yet)
\footnote{The form $\exp \Bigl({ -s \tilde
    \gamma_+(1+\delta)/(2\beta_0)} \Bigr)$ coincides with the both solution:
  Eq.(\ref{9}) if $x^{\hat d_+} >>1$ and Eq.(\ref{11}) when $\delta
  =0$ but it is not the solution of DGLAP equation.}. We
introduce the some ``critical'' value of $Q^2$: $Q^2_c$, where the
solution (\ref{9}) is replaced by the solution (\ref{11}). The
exact value of $Q^2_c$ may be obtained from the fit of experimental
data. Thus, we have in the LO of the perturbation theory:

 \begin{eqnarray}
\z f_a(x,t)~=~ f_a^-(x,t)~+~ f_a^+(x,t) \nonumber  \\
\z f_a^-(x,t)~=~A^-_a~\exp{(- d_-s)} \nonumber  \\
% \label{12}  \\
%\[
\z f_a^+(x,t)~=~
%A^+_a~
\left\{
\begin{array}{ll} A^+_a
x^{\hat d_+s}\exp{(-\overline d_+s)}, & \mbox{ if } Q^2 \leq Q^2_c \\
%x^{-\hat d_+s_c}e^{-\overline d_+s_c}
f_a^+(x,t_c)
\exp{\Bigl(-d_+(1+\delta_c)(s-s_c)\Bigr)},
& \mbox{ if } Q^2>Q^2_c
\end{array} \right.
%\]
 \label{12}
 \end{eqnarray}
where
%(see, for example, \cite{14})

 \begin{eqnarray}
\z t_c~=~t(Q^2_c),~~s_c~=~s(Q^2_c) \nonumber  \\
\z
A^+_q~=~(1- \overline \alpha )A_q ~+~ \tilde \alpha A_g,~~
A^+_g~=~ \overline \alpha A_g ~-~ \varepsilon A_q \nonumber \\
\z\mbox{and } A_a^-~=~A_a ~-~ A_a^+
 \label{13}
 \end{eqnarray}
and the values of the coefficients $\overline \alpha$, $\tilde \alpha$
and $\varepsilon$ may be found, for example, in \cite{18}.

Using the concrete AD values at $\delta =0$ and $f=3$, we have

 \begin{eqnarray} \z
A^+_q~ \approx ~\frac{1}{27}\frac{4A_q+9A_g}{ln(\frac{1}{x})-\varrho
  (\nu) - \frac{85}{108}}  \nonumber \\
\z A^+_g~ \approx ~A_g~+~\frac{4}{9}A_q ~-~
\frac{4}{27}\frac{9A_g-A_q}{ln(\frac{1}{x})-\varrho
  (\nu) - \frac{85}{108}}
 \label{14}
 \end{eqnarray}

Thus, the value of the ``+''component of the quark PD is suppressed
logarithmically
that is in the qualitative agreement with the $HERA$ parametrizations of
SF $F_2$ (see \cite{20.5,21})
(in the LO $F_2(x,Q^2)~=~(2/9)f_q(x,Q^2)$ for $f=3$), where
the magnitude connected
with the factor $x^{-\delta}$ is $5 \div 10 \%$ from the flat (for $x \to
                                        0$) magnitude.

{\bf 2.} By analogy with the subsection {\bf 1} and knowing the NLO
$Q^2$-dependence of PD moments, we obtain the following equations for the NLO
$Q^2$-evolution of the both: ''+'' and ``$-$'' PD components (hereafter
$\tilde s=ln(\alpha(Q^2_0)/\alpha(Q^2)), p=\alpha(Q^2_0)-\alpha(Q^2)$):

 \begin{eqnarray}
\z f_a(x,t)~=~ f_a^-(x,t)~+~ f_a^+(x,t) \nonumber  \\
\z f_a^-(x,t)=~\tilde A^-_a~\exp{(- d_-\tilde s -d_{--}^ap)}
 \nonumber  \\
%\[
\z f_a^+(x,t)=
%\tilde A^+_a~
\left\{
\begin{array}{ll} \tilde A^+_a
x^{(\hat d_+\tilde s + \hat d_{++}^a p)}\exp{(-\overline d_+\tilde s
  -\overline d_{++}^ap)}, & \mbox{if } Q^2 \leq Q^2_c \\
%x^{-(\hat d_+\tilde s_c+ \hat d_{++}p_c)}e^{-(\overline d_+\tilde s_c+
%  \overline d_{++}p_c}
f_a^+(x,t_c)
\exp{\Bigl(-d_+(1+\delta_c)(\tilde s-\tilde
  s_c)-d_{++}^a(1+\delta_c)(p-p_c) \Bigr) },
& \mbox{if } Q^2>Q^2_c
\end{array} \right.
%\]
 \label{15}
 \end{eqnarray}
where
%(see, for example, \cite{14})

 \begin{eqnarray} \z
\tilde s_c ~=~ \tilde s(Q^2_c),~
p_c~=~p(Q^2_c),~\alpha_0~=~\alpha(Q^2_0)
,~\alpha_c~=~\alpha(Q^2_c) \nonumber \\ \z
\tilde A^{\pm}_a~=~\Bigl(1~-~\alpha_0 K^a_{\pm} \Bigr)
A^{\pm}_a ~+~ \alpha_0 K^a_{\pm} A^{\mp}_a \nonumber \\  \z
d_{++}^a~=~ \hat d_{++}^a
\Bigl( ln(\frac{1}{x}) - \varrho(\nu) \Bigl) ~+~ \overline d_{++}^a, ~~
d^a_{++}~=~ \frac{\gamma_{\pm \pm}}{2\beta_0} ~-~
\frac{\gamma_{\pm} \beta_1}{2\beta^2_0} ~-~ K^a_{\pm}  \nonumber \\
\z\mbox{and }~~ K^q_{\pm}~=~ \frac{\gamma_{\pm \mp}}{2\beta_0 +
  \gamma_{\pm} - \gamma_{\mp}},~~ K^g_{\pm}~=~ K^q_{\pm}
\frac{\gamma_{\pm}-\gamma^{(0)}_{qq}}{\gamma_{\mp}-\gamma^{(0)}_{qq}}
 \label{16}
 \end{eqnarray}

The NLO AD of the ``$\pm$'' components are connected with the NLO AD
$\gamma^{(1)}_{ab}$. The corresponding formulae can be found in
\cite{18}.

Using the concrete values of the LO and  NLO AD at $\delta =0$ and
$f=3$, we obtain the following values for the NLO components from
(\ref{15}),(\ref{16}) (note that we remail only the terms $\sim O(1)$
in the NLO terms)

 \begin{eqnarray} \z
 d^q_{--}~=~ \frac{16}{81} \Big[ 2\zeta (3) + 9 \zeta (2) -
\frac{779}{108}  \Big] \approx 1.97, ~~
d^g_{--}~=~ d^q_{--}~+~ \frac{28}{81} \approx 2.32 \nonumber \\ \z
\hat d^q_{++}~=~ \frac{2800}{81} , ~~
\overline d^q_{++}~=~ 32 \Big[ \zeta (3) + \frac{263}{216}\zeta (2) -
\frac{372607}{69984}  \Big] \approx -67.82    \nonumber \\ \z
\hat d^g_{++}~=~ \frac{1180}{81} , ~~
\overline d^g_{++}~=~ \overline d^q_{++}~+~ \frac{953}{27} -12\zeta
(2) \approx -52.26
 \label{17}
 \end{eqnarray}
and

 \begin{eqnarray} \z
\tilde A^+_q~ \simeq ~\frac{20}{3} \alpha_0
\Bigl[ A_g + \frac{4}{9} A_q \Bigr] ~+~
\frac{1}{27}\frac{4A_q(1-7.67 \alpha_0)+9A_g(1-8.71
  \alpha_0)}{ln(\frac{1}{x})-\varrho
  (\nu) - \frac{85}{108}}  \nonumber \\
\z \tilde A^+_g~ \simeq ~ \Bigl(A_g ~+~\frac{4}{9}A_q \Bigr)
 \Bigl(1-\frac{80}{9}\alpha_0 \Bigr)
 ~-~
\frac{4}{27}\frac{9A_g-A_q}{ln(\frac{1}{x})-\varrho
  (\nu) - \frac{85}{108}} \Bigl( 1+ \frac{692}{81}\alpha_0)
 \nonumber \\
\z\mbox{and } \tilde A_a^-~=~A_a ~-~ \tilde A_a^+
 \label{18}
 \end{eqnarray}

It is useful to change  in Eqs.(\ref{15})-(\ref{18}) from the quark
PD to the SF $F_2(x,Q^2)$, which is connected in NLO approximation with the PD
by the following way (see \cite{18}):

 \begin{eqnarray}
F_2(x,Q^2)~=~ \Bigl(  1+\alpha(Q^2)B_q(1+\delta) \Bigr) \delta^2_s
f_q(x,Q^2) ~+~ \alpha(Q^2)B_g(1+\delta) \delta^2_s f_g(x,Q^2),
 \label{19}
 \end{eqnarray}
where $\delta ^2_s = \sum_{i=1}^f/f \equiv <e_f^2>$ is the average
charge square of the active quarks: $\delta ^2_s$ = (2/9 and 5/18) for
$f$ = (3 and 4), respectively.
 The NLO corrections lead to the
appearence in the r.h.s. of Eqs.(\ref{15}) of the additional terms
$\Bigl(  1+\alpha B_{\pm} \Bigr)/\Bigl(  1+\alpha_0 B_{\pm} \Bigr)$
and the necessarity to transform
$\tilde A^{\pm}_q$ to $C^{\pm} \equiv F_2^{\pm}(x,Q^2)$ into the input
parts. The final results for $F_2(x,Q^2)$ are in the form:

 \begin{eqnarray}
\z F_2(x,t)~=~ F_2^-(x,t)~+~ F_2^+(x,t) \nonumber  \\
\z F_2^-(x,t)~=~ C^-~\exp{(- d_-\tilde s -d_{--}^qp)}
(1+\alpha B^-)/(1+\alpha_0 B^-)
 \nonumber  \\
% \label{16}  \\
%\[
\z F_2^+(x,t)~=~
%\tilde A^+_a~
\left\{
\begin{array}{ll} C^+
x^{(\hat d_+\tilde s + \hat d_{++}^q p)}\exp{(-\overline d_+\tilde s
  -\overline d_{++}^qp)}(1+\alpha B^+)/(1+\alpha_0 B^+)
, & \mbox{ if } Q^2 \leq Q^2_c \\
%x^{-(\hat d_+\tilde s_c+ \hat d_{++}p_c)}e^{-(\overline d_+\tilde s_c+
%  \overline d_{++}p_c}
F_2^+(x,t_c)
\exp{\Bigl(-d_+(1+\delta_c)(\tilde s-\tilde
  s_c)-d_{++}^q(1+\delta_c)(p-p_c) \Bigr) } &  \\
\biggl(1+
%\Bigl[\alpha(Q^2)-\alpha(Q^2_c) \Bigr]
\alpha B^+(1+\delta_c) \biggr)/
\biggl(1+
%\Bigl[\alpha(Q^2)-\alpha(Q^2_c) \Bigr]
\alpha_c B^+(1+\delta_c) \biggr),
& \mbox{ if } Q^2>Q^2_c
\end{array} \right.
%\]
 \label{20}
 \end{eqnarray}
where
%(see, for example, \cite{14})
$$ B^{\pm}~=~B_q ~+~ \frac{\gamma_{\pm}}{\gamma^{(0)}_{qg}}B_g,~~
C^{\pm}~=~\tilde A^{\pm}_q (1+\alpha_0 B^{\pm})$$
with the substitution of $A_q$ by $C \equiv F_2(x,Q^2_0)$ into
Eq.(\ref{18}) $\tilde
A^{\pm}_q$ according

 \begin{eqnarray} \z
C~=~ \Bigl(  1+\alpha_0 B_q \Bigr) \delta^2_s
A_q ~+~ \alpha_0 B_g \delta^2_s A_g,
 \label{21}
 \end{eqnarray}

For the gluon PD the situation is more simple: in
Eq.(\ref{18}) it is necessary to replace $A_q$ by $C$
according (\ref{21}).

For the concrete values of the LO and NLO AD at $\delta =0$ and $f=3$,
we have for $Q^2$-evolution of $F_2(x,Q^2)$ and the gluon PD:

 \begin{eqnarray}
\z F_2(x,t)~=~ F_2^-(x,t)~+~ F_2^+(x,t),~~
f_g(x,t)~=~ f_g^-(x,t)~+~ f_g^+(x,t) \nonumber  \\
\z F_2^-(x,t)~=~ C^-~\exp{(- \frac{32}{81} \tilde s
  -1.97p)}(1-\frac{8}{9} \alpha )/(1-\frac{8}{9} \alpha_0 )
 \nonumber  \\
% \label{16}  \\
%\[
\z F_2^+(x,t)~=~
%\tilde A^+_a~
\left\{
\begin{array}{ll} C^+
x^{(-\frac{4}{3} \tilde s + \frac{2800}{81}
  p)}
\exp{\Bigl(- \frac{4}{3}(\varrho(\nu)+\frac{101}{108}) \tilde s
  +(\frac{2800}{81} \varrho(\nu)-67.82)p \Bigr)}  &  \\
\Bigl(1+6[ln(\frac{1}{x})-\varrho(\nu)-\frac{101}{108}] \alpha \Bigr)/
\Bigl(1+6[ln(\frac{1}{x})-\varrho(\nu)-\frac{101}{108}] \alpha_0 \Bigr)
, & \mbox{if } Q^2 \leq Q^2_c \\
%x^{-(\hat d_+\tilde s_c+ \hat d_{++}p_c)}e^{-(\overline d_+\tilde s_c+
%  \overline d_{++}p_c}
F_2^+(x,t_c)
\exp{\Bigl(-d_+(1+\delta_c)(\tilde s-\tilde
  s_c)-d_{++}^q(1+\delta_c)(p-p_c) \Bigr) }  &  \\
\biggl(1+
%\Bigl[\alpha(Q^2)-\alpha(Q^2_c) \Bigr]
\alpha B^+(1+\delta_c) \biggr)/
\biggl(1+
%\Bigl[\alpha(Q^2)-\alpha(Q^2_c) \Bigr]
 \alpha_c B^+(1+\delta_c) \biggr),
& \mbox{if } Q^2>Q^2_c
\end{array} \right.
%\]
 \label{22}  \\
\z f_g^-(x,t)~=~ A_g^-~\exp{(- \frac{32}{81} \tilde s
  -2.32p)}(1-\frac{8}{9} \alpha )/(1-\frac{8}{9} \alpha_0 )
 \nonumber  \\
% \label{16}  \\
%\[
\z f_g^+(x,t)~=~
%\tilde A^+_a~
\left\{
\begin{array}{ll} A_g^+
x^{(-\frac{4}{3} \tilde s + \frac{1180}{81}
  p)}\exp{\Bigl(- \frac{4}{3}(\varrho(\nu)+\frac{101}{108}) \tilde s
  +(\frac{1180}{81} \varrho(\nu)-52.26)p \Bigr)} &  \\
\Bigl(1+6[ln(\frac{1}{x})-\varrho(\nu)-\frac{101}{108}] \alpha \Bigr)/
\Bigl(1+6[ln(\frac{1}{x})-\varrho(\nu)-\frac{101}{108}] \alpha_0 \Bigr)
, & \mbox{if } Q^2 \leq Q^2_c \\
%x^{-(\hat d_+\tilde s_c+ \hat d_{++}p_c)}e^{-(\overline d_+\tilde s_c+
%  \overline d_{++}p_c}
f_g^+(x,t_c)
\exp{\Bigl(-d_+(1+\delta_c)(\tilde s-\tilde
  s_c)+d_{++}^a(1+\delta_c)(p-p_c) \Bigr) } &   \\
\biggl(1+
%\Bigl[\alpha(Q^2)-\alpha(Q^2_c) \Bigr]
\alpha B^+(1+\delta_c) \biggr)/
\biggl(1+
%\Bigl[\alpha(Q^2)-\alpha(Q^2_c) \Bigr]
 \alpha_c B^+(1+\delta_c) \biggr),
& \mbox{ if } Q^2>Q^2_c
\end{array} \right.
%\]
 \label{23}
 \end{eqnarray}
where

 \begin{eqnarray} \z
\tilde C^+~ \simeq ~\frac{2}{27}
\Biggl(  26\alpha_0
\Bigl[ A_g + 2C \Bigr] ~+~
\frac{A_g(1-9.74 \alpha_0)+2C(1-7.82
  \alpha_0)}{ln(\frac{1}{x})-\varrho
  (\nu) - \frac{85}{108}} \Biggr) \nonumber \\
\z\mbox{and }  C^-~=~C
%~-~ \frac{4}{9}\alpha_0 A_g
{}~-~ C^+
 \label{24}  \\
\z \tilde A^+_g~ \simeq ~A_g \Bigl(1-\frac{28}{3}\alpha_0 \Bigr)
{}~+~2C ~-~
\frac{2}{27}\frac{2A_g(1+ \frac{590}{81}\alpha_0)-C(1+
\frac{572}{81}\alpha_0)}{ln(\frac{1}{x})-\varrho
  (\nu) - \frac{85}{108}}
 \nonumber \\
\z\mbox{and } \tilde A_g^-~=~A_g ~-~ \tilde A_g^+
 \label{25}
 \end{eqnarray}

Let us give some conclusions following from
Eqs.(\ref{24})-(\ref{25}). It is clearly seen that the NLO
corrections reduce the LO contributions. Indeed, the value of the
subcritical Pomeron intercept, which increases as $ln(\alpha_0/\alpha)$ in the
LO,  obtaines the additional  term $ \sim (\alpha_0 - \alpha)$ with the
large (and opposite in sign to the LO term) numerical coefficient. Note
that this coefficient is different for the quark and gluon PD, that is
in the agreement with the recent $MRS(G)$ fit in \cite{19} and the
data analysis by $ZEUS$ group (see \cite{20}). The intercept of the
gluon PD is larger then the quark PD one (see also \cite{19,20}).
However, the effective reduction of the quark PD is smaller (that is in
the agreement with W.-K. Tung analysis in \cite{21}), because the quark PD
 part increasing at small $x$  obtains the additional ($ \sim
\alpha_0$ but not $ \sim 1/lnx $) term, which is  important at very small
$x$.

Note that there is the fourth quark threshold at $Q^2_{th} \sim 10
GeV^2$ and the $Q^2_{th}$ value may be larger or smaller to $Q^2_c$
one. Then, either the solution in the r.h.s. of
Eqs. (\ref{20},\ref{22},\ref{23})
before the critical point $Q^2_c$ and the one for $Q^2 > Q^2_c$ contain
the threshold transition, where the values of all variables are
changed from
ones at $f=3$ to ones at $f=4$. The $\alpha(Q^2)$ is smooth because
$\Lambda^{f=3}_{\overline{MS}} \to  \Lambda^{f=4}_{\overline{MS}}$
(see also the recent experimental test of the flavour independence of
strong interactions into \cite{22}).

For simplicity here we suppose that $Q^2_{th} = Q^2_c$ and all changes
initiated by threshold are done authomatically: the first (at $Q^2
\leq Q^2_c$) solutions contain $f=3$ and second (at $Q^2 > Q^2_c$)
ones have $f=4$,
respectively. For the ``$-$'' component we should use $Q^2_{th}=Q^2_c$,
too.

Note only that the Pomeron intercept $\alpha_p = 1~-~(d_+ \tilde s +
\hat d^q_{++}p)$ increases at $Q^2=Q^2_{th}$, because

\[
\alpha_p ~-~1 ~=~
\left\{
\begin{array}{ll}
\frac{4}{3} \tilde s(Q^2_{th},Q^2_0)~-~ \frac{2800}{81} p(Q^2_{th},Q^2_0)
 , & \mbox{ if } Q^2 \leq Q^2_c \\
1.44 \tilde s(Q^2_{th},Q^2_0)~-~ 38.11 p(Q^2_{th},Q^2_0)
,& \mbox{ if } Q^2>Q^2_c
\end{array} \right.
\]
that agrees
%in principle
with results \cite{23} obtained in
the framework of dual parton model. The difference
%in the pomeron intercept
$$ \bigtriangleup \alpha_p ~=~ 0.11 \tilde s(Q^2_{th},Q^2_0) - 3.55
p(Q^2_{th},Q^2_0) $$
dependes from the values of $Q^2_{th}$ and $Q^2_0$.
 For $Q^2_{th}=10GeV^2$ and $Q^2_0=1GeV^2$ it is very small:
$$ \bigtriangleup \alpha_p ~=~ 0.012 $$

{\bf 3.}
Let us resume the obtained results. We have got the DGLAP equation
``solution'' having the Regge form (\ref{1}) for the two cases: at small
$Q^2$ ($Q^2 \sim 1GeV^2$), where SF and PD have the flat behaviour at
small $x$, and at large $Q^2$, where SF $F_2(x,Q^2)$ fastly increases
when $x \to 0$. The behaviour in the flat case is nonstable with the
perturbative viewpoint because it leads to the production of the
subcritical value of pomeron intercept at larger $Q^2$ and the its
increase (like $4/3~ ln(\alpha (Q^2_0)/\alpha(Q^2)$ in LO) when the
 $Q^2$ value increases\footnote{The Pomeron intercept value increasing with
  $Q^2$ was obtained also in \cite{23.5}.}. The solution in the Lipatov
Pomeron case corresponds to the well-known results (see
\cite{12,14,17}) with $Q^2$-independent Pomeron intercept. The general
``solution'' should contains the smooth transition between these
pictures. Unfortunately, it is impossible to obtain it in the case of
the simple approximation (\ref{1}), because the r.h.s. of DGLAP
equation (\ref{7}) contains the both: $\sim x^{-\delta}$ and $\sim
Const$, terms. As a result, we used two above ``solutions'' gluing in
some point $Q^2_c$.

Note that our ``solution'' is some generation (or a application) of
the solution of DGLAP equation in the momentum space. The last
one have two: ''+'' and ``$-$'' components. The above our
conclusions are related to the ``+'' component, which is the basic
Regge asymptotic. The Pomeron intercept corresponding to ``$-$''
component, is $Q^2$-independent and this component is the
subasymptotical one at large $Q^2$. However, the magnitude
of the ``+'' is suppressed
like $1/ln(1/x)$ and $\alpha (Q^2_0)$, and the subasymptotical ``$-$''
component may be important. Indeed, it is observed experimentally (see
\cite{20.5,20}). Note, however, that the suppression $\sim
\alpha(Q^2_0)$ is really very slight if we choose a small value of
$Q^2_0$.

Our ``solution'' in the form of Eqs.(\ref{22})-(\ref{25}) is in the very
well agreement with the recent $MRS(G)$ fit \cite{19} and with the
results of \cite{17} at $Q^2=15GeV^2$. As it can be seen from
Eqs.(\ref{22}),(\ref{23}), in our formulae there is the dependence
on the PD behaviour
at large $x$. Following to \cite{25} we choose $\nu =5$ that
agrees in the
gluon case with the quark counting rule \cite{26}.
This $\nu$ value is also close to  the values obtained by
$CCFR$ group \cite{27} ($\nu = 4$) and in the last $MRS(G)$ analysis
\cite{19} ($\nu =6$). Note that this dependence is strongly reduced
for the gluon PD in the form

$$ f_g(x,Q^2_0)~=~A_g(\nu)(1-x)^{\nu}, $$
if we suppose that the proton's momentum is carred by gluon, is
 $\nu$-independent. We used $A_g(5)=2.1$ and $F_2(x,Q^2_0)=0.3$ when
$x \to 0$.

For the quark PD the choise $\nu =3$ is more preferable, however the
use of  two different $\nu$ values complicates the analysis. Because
the quark contribution to the ``+'' component is not large, we put
$\nu =5$ to both: quark and gluon cases. Note also that the variable
$\nu (Q^2)$ have (see \cite{28}) the $Q^2$-dependence determinated by
the LO AD $\gamma^{(0)}_{NS}$. However this $Q^2$-dependence is
proportional $s$ and it is not important in our analysis.

Starting from $Q^2_0=1GeV^2$ (by analogy with \cite{23.6}) and from
$Q^2_0=2GeV^2$, and using two values of QCD parameter $\Lambda$: more
standard one ($\Lambda^{f=4}_{\overline {MS}}$ = 200 $MeV$) and
 ($\Lambda^{f=4}_{\overline {MS}}$= 255 $MeV$) obtained in \cite{19},
we have the following values of the quark and gluon PD ``intercepts''
$\delta_a ~=~-(d_+ \tilde s +\hat d^q_{++}a)$ (here
$\Lambda^{f=4}_{\overline {MS}}$ is marked as $\Lambda$):

if $Q^2_0$ = 1 $GeV^2$

\begin{center}
\begin{tabular}{|l||l|l|l|l|}         \hline
$Q^2$ & $\delta_q(Q^2)$  & $ \delta_g(Q^2)$ & $\delta_q(Q^2)$  &
$\delta_g(Q^2)$ \\
    &$\Lambda =200MeV$ &$\Lambda =200MeV$ &$\Lambda =255MeV$ &$\Lambda
    =255MeV$ \\   \hline
4   &   0.191        &   0.389        &   0.165        &   0.447   \\ \hline
10  &   0.318        &   0.583        &   0.295        &   0.659   \\ \hline
15  &   0.367        &   0.652        &   0.345        &   0.734   \\ \hline
\end{tabular}
\end{center}

if $Q^2_0$ = 2 $GeV^2$

\begin{center}
\begin{tabular}{|l||l|l|l|l|}         \hline
$Q^2$ & $\delta_q(Q^2)$  & $ \delta_g(Q^2)$ & $\delta_q(Q^2)$  &
$\delta_g(Q^2)$ \\
    &$\Lambda =200MeV$ &$\Lambda =200MeV$ &$\Lambda =255MeV$ &$\Lambda
    =255MeV$ \\   \hline
4   &   0.099        &   0.175        &   0.097        &   0.198   \\ \hline
10  &   0.226        &   0.368        &   0.227        &   0.410   \\ \hline
15  &   0.275        &   0.438        &   0.278        &   0.486   \\ \hline
\end{tabular}
\end{center}

Note that these values of $\delta_a $ are above the ones from
\cite{19}. Because we have the second (subasymptotical) part, the
effective our ``intercepts'' have the smaller values.

As a conclusion, we note that BFKL equation (and thus the value of
Lipatov Pomeron intercept) was obtained in \cite{19.5} in the
framework of perturbative QCD. The large-$Q^2$ $HERA$ experimental
data are in the good agreement with Lipatov's trajectory and thus with
perturbative QCD. The small $Q^2$ data agrees with the standard
Pomeron intercept $\alpha_p=1$ or with Donnachie-Landshoff pisture:
$\alpha_p=1.08$. Perhaps, this range requires already the knowledge of
nonperturbative QCD dynamics and  perturbative solutions (including
BFKL one) should be not applied here directly and are corrected by some
 nonperturbative contributions.

In our analysis Eq.(\ref{1}) can be considered as the nonperturbative
(Regge-type) input at $Q^2_0 \sim 1GeV^2$. Above $Q^2_0$ the PD
behaviour obeys DGLAP equation, Pomeron moves to the subcritical
regime and tends to its perturbative value. After some $Q^2_c$, where
its perturbative value was already  attained, Pomeron intercept saves
the permanent value. The application of this approach to analyse small
$x$ data invites futher investigation.

%\footnote{\footnote{\footnote{

\end{document}